
\input amstex
\magnification 1200
\documentstyle{amsppt}
\NoBlackBoxes
\NoRunningHeads
\def\L{\Lambda}
\def\e{\varepsilon}

\def\l{\lambda}

\def\Z{\Bbb Z}
\def\C{\Bbb C}

\def\U{U_q(\widehat{\frak{sl}_2})}
\def\Uf{U_q({\frak{sl}_2})}

\def\d{\partial}

\def\<{\langle}
\def\>{\rangle}
\def\o{\otimes}

\topmatter
\title On spectral theory of quantum vertex operators
\endtitle
\author {\rm {\bf Pavel I. Etingof} \linebreak
\vskip .1in
Department of Mathematics\linebreak
Harvard University\linebreak
Cambridge, MA 02138, USA\linebreak
e-mail: etingof\@math.harvard.edu}
\endauthor

\endtopmatter
\centerline{October 27, 1994}
\vskip .1in

In this note we prove a conjecture from \cite{DFJMN} on the asymptotics
of the composition of $n$ quantum vertex operators for the quantum affine
algebra $\U$, as $n$ goes to $\infty$.
For this purpose we define and study the leading eigenvalue and
eigenvector of the product of two components of the quantum vertex operator.
This eigenvector and the corresponding eigenvalue
were recently computed by M.Jimbo. The results of his computation are given in
Section 4.

\vskip .1in
\centerline{\bf 1. Basic definitions}

\vskip .1in
1.1. {\it Quantum groups. }
Let $\U$ be the quantum group generated over $\C(q)$
by the elements $e,f,t^{\pm 1}$
satisfying the standard relations:
$$
tet^{-1}=q^2e, tft^{-1}=q^{-2}f, [e,f]=\frac{t-t^{-1}}{q-q^{-1}}.\tag 1.1
$$

For an integer $n$, set $[n]=\frac{q^n-q^{-n}}{q-q^{-1}}$.

Let $\U$ be the quantum affine algebra generated over $\C(q)$ by
the elements $e_i,f_i,t_i^{\pm 1}$, $i=0,1$
satisfying the standard relations:
$$
\gather
t_ie_it_i^{-1}=q^2e_i, t_if_it_i^{-1}=q^{-2}f_i, [e_i,f_i]=
\frac{t_i-t_i^{-1}}{q-q^{-1}}, i=0,1;\\
[e_i,f_j]=0,
e_i^3e_j-[3]e_i^2e_je_i+[3]e_ie_je_i^2-e_je_i^3=0,\\
f_i^3f_j-[3]f_i^2f_jf_i+[3]f_if_jf_i^2-f_jf_i^3=0, i\ne j\\
.\tag 1.2
\endgather
$$
We define the coproduct by
$\Delta(t_i)=t_i\o t_i,\Delta(e_i)=e_i\o 1+t_i\o e_i,\Delta(f_i)=
f_i\o t_i^{-1}+1\o f_i$. Tensor product of representations of $\U$ is defined
with the help of this coproduct.

For $z\in\C^*$, let $p_z:\U\to \Uf$ be the
evaluation homomorphism defined by $e_0\to zf,
f_0\to z^{-1}e, t_0\to t^{-1}, e_1\to e, f_1\to f, t_1\to t$.

\vskip .1in

1.2. {\it Representations. }
Let $\Lambda_0$, $\Lambda_1$ be the fundamental weights for the $\U$.
Let $L_0=V(\L_0)$, $L_1=V(\L_1)$
denote the irreducible integrable highest weight
representations of the quantum affine algebra $\U$ with highest
weights $\Lambda_0,\Lambda_1$, respectively.
Let $v_0,v_1$ be their highest weight vectors.
Let $L=L_0\oplus L_1$. Let $\hat L$, $\hat L_i$ be the completions of the
modules $L,L_i$ with respect to the homogeneous grading.

 Let $V$ be the two-dimensional irreducible representation of $\Uf$
in which the spectrum of $t$ is $q,q^{-1}$. Let $v_+,v_-$ be a basis of this
representation such that $tv_{\pm}=q^{\pm 1}v_{\pm}$, and $v_-=fv_+$.
Let $V(z)=p_z^*V$ be the representation of $\U$ obtained by pullback of $V$
by $p_z$.

\vskip .1in

1.3. {\it Vertex operators.} Quantum vertex operators were introduced by
I.Frenkel and N.Reshetikhin.
It is known \cite{DJO}
that for any $z\in\C^*$
there exist unique intertwining operators
$$
\Phi^0(z): L_0\to \hat L_1\o V(z), \Phi^1(z): L_1\to \hat L_0\o V(z),\tag 1.3
$$
such that $\Phi^0(z)v_0=v_1\o v_-+ \text{lower weight terms}$,
$\Phi^1(z)v_1=v_0\o v_++ \text{lower weight terms}$ (by ``weight'' we mean
the weight of the first component). These operators are called quantum vertex
operators. Let $\Phi(z) :L\to \hat L\o V$ be defined by
$\Phi=\Phi^0\oplus\Phi^1$. We define the operators
$\Phi_{\pm}(z):L\to\hat L$ by
$$
\Phi(z)=\Phi_+(z)\o v_++\Phi_-(z)\o v_-.\tag 1.4
$$
It is easy to see that $t\Phi_{\pm}t^{-1}=q^{\mp 1}\Phi_{\pm}$.

\vskip .1in

1.4. {\it The Fock space. } We would like to study the dependence
of vertex operators on the parameter $q$. For this purpose we will
1) realize the $\C(q)$-vector
space $L$ as $\C(q)\o_\C H$, where
$H$ is a complex vector space called the Fock space,
 and 2) write down the action
of the quantum group and
vertex operators in $\C(q)\o H$ as series in $q$ whose coefficients
are operators on $H$. This construction is called bosonization
and comes from \cite{FJ,JMMN}.

Let us now define the Fock space $H$. Let $\frak h$ be the Heisenberg
Lie algebra with the basis $\{b_i, i\in\Z\setminus \{0\};Z\}$, and
relations
$$
[b_m,b_n]=m\delta_{m+n,0}Z;\ [X,Z]=0, X\in\frak h.\tag 1.5
$$
Let $H_0=\C[b_{-1},b_{-2},...]$. Then $H_0$ is naturally a
representation of $\frak h$, in which $Z=1$, and
$b_n$ acts by multiplication by
itself for $n<0$, and by differentiation $n\frac{\d}{\d b_{-n}}$ for
$n>0$. Let $H=H_0\o \C[\Z]$.

We denote the element of $\C[\Z]$ corresponding to the integer $n$
by $\e^n$. We introduce the homogeneous gradation in $H$ in a standard
way: the degree of $b_{-n}$ is $-n$, $n>0$, and the degree of $\e^n$ is
$(i-n^2)/4$, where $i=1$ if $n$ is odd and $0$ if $n$ is even.

\vskip .1in

1.5. {\it Bosonization of $\U$. }
Now let us define the action of $\U$ in $H$.
Set
$$
a_n=q^{-n/2}\frac{[n]}{n}b_n,
\ a_{-n}=q^{n/2}\frac{[2n]}{n}b_{-n},\ n>0.\tag 1.6
$$
Then we have
$$
[a_m,a_n]=\delta_{m+n,0}\frac{[m][2m]}{m}Z.\tag 1.7
$$

Let
$$
\gather
X^{\pm}(z)=\sum_{n\in\Z}X^{\pm}_nz^{-n-1}=\\
\exp\biggl(\pm\sum_{n=1}^{\infty}\frac{a_{-n}}{[n]}q^{\mp n/2}z^n\biggr)
\exp\biggl(\mp\sum_{n=1}^{\infty}\frac{a_{n}}{[n]}q^{\mp n/2}z^{-n}\biggr)
\o\e^{\pm 2}z^{\pm \d_\e},\tag 1.8\endgather
$$
where the first component acts in $H_0$, the second component acts in $\C[\Z]$,
and $\d_\e$ is defined by $\d_\e \e^n=n\e^n$. Then all Fourier
coefficients of this series define linear operators on the space
$\C(q^{1/2})\o H$.

\proclaim{Theorem 1.1} (I.Frenkel-N.Jing,\cite{FJ}) There exists a unique
representation of $\U$ in $\C(q^{1/2})\o H$ such that
$$
\gather
t_1\to 1\o q^{\d_\e}, t_0\to 1\o q^{1-\d_\e}, e_1\to X^+_0, f_1\to X^-_0,\\
e_0\to X_1^-(1\o q^{-\d_\e}), f_0\to (1\o q^{1-\d_\e})X_{-1}^+.
\tag 1.9\endgather
$$
This representation is isomorphic to $\C(q^{1/2})\o_{\C(q)} L$.
The gradation in
$\C(q^{1/2})\o H$ introduced above
coincides with the homogeneous gradation in $L$.
\endproclaim

Let us rewrite (1.8) in terms of $\{b_n\}$:
$$
\gather
X^+(z)=
\exp\biggl(\sum_{n=1}^{\infty}\frac{b_{-n}}{n}(q^n+q^{-n})z^n\biggr)
\exp\biggl(-\sum_{n=1}^{\infty}\frac{b_{n}}{n}q^{-n}z^{-n}\biggr)
\o\e^{ 2}z^{ \d_\e},\\
X^-(z)=
\exp\biggl(-\sum_{n=1}^{\infty}\frac{b_{-n}}{n}(q^{2n}+1)z^n\biggr)
\exp\biggl(\sum_{n=1}^{\infty}\frac{b_{n}}{n}z^{-n}\biggr)
\o\e^{-2}z^{- \d_\e}.\tag 1.10\endgather
$$
It is seen from this equation that
in fact, the representation of $\U$ defined by (1.9) is well defined over
$\C(q)$ if considered in the basis of polynomials of $b_{-n}$
(it is not necessary to take the square root of $q$).
We can also see from (1.10)
that $X^-_n$ are actually defined over polynomials in $q$.
This fact will be used later.

 From now on we identify $\C(q)\o H$ and $L$ by the $\U$-isomorphism
$\C(q)\o H\to L$ fixed by the conditions
$1\o\e^0\to v_0, 1\o\e^1\to v_1$.

\vskip .1in

1.6. {\it Bosonization of vertex operators.}

Let $I: \C[\Z]\to\C[\Z]$ be defined by $I\e^n=\frac{1}{2}(1-(-1)^n)\e^n$.

\proclaim{Theorem 1.2} (\cite{JMMN}) The vertex operators
$\Phi_{\pm}(z): L\to \hat L$ are given by the formulas
$$
\gather
\Phi_-(z)=
\exp\biggl(\sum_{n=1}^{\infty}\frac{a_{-n}}{[2n]}q^{7n/2}z^n\biggr)
\exp\biggl(-\sum_{n=1}^{\infty}\frac{a_{n}}{[2n]}q^{-5n/2}z^{-n}\biggr)
\o\e^{1}(-q^3z)^{(\d_\e+I)/2},\\
\Phi_+(z)=\Phi_-(z)X_0^--qX_0^-\Phi_-(z).\tag 1.11\endgather
$$
\endproclaim

Let us write down
the expression of
the vertex operators
in terms of $\{b_n\}$. We have
$$
\Phi_-(-q^{-3}z)=
\exp\biggl(\sum_{n=1}^{\infty}\frac{(-1)^nq^nb_{-n}}{n}z^n\biggr)
\exp\biggl(-\sum_{n=1}^{\infty}\frac{(-1)^nq^nb_{n}}{n(1+q^{2n})}z^{-n}\biggr)
\o\e^{1}z^{(\d_\e+I)/2}.\tag 1.12
$$
This shows, in particular, that
 we do not in fact need $q^{1/2}$, i.e everything is defined over $\C(q)$.

\vskip .1in

1.7. {\it Boson-Fermion correspondence.}
Boson-Fermion correspondence was first discussed in physics literature
\cite{BH}. A representation-theoretic
 description of this correspondence is given in \cite{F}.

Consider the following formal series in $z$:
$$
\gather
\psi(z)=\sum_{n\in\Z}\psi_nz^{-n}=\\
\exp\biggl(\sum_{n=1}^{\infty}\frac{b_{-n}}{n}z^n\biggr)
\exp\biggl(-\sum_{n=1}^{\infty}\frac{b_{n}}{n}z^{-n}\biggr)\o \e^{1}
z^{\d_\e+1},\\
\psi^*(z)=\sum_{n\in\Z}\psi^*_nz^{-n}=\\
\exp\biggl(-\sum_{n=1}^{\infty}\frac{b_{-n}}{n}z^n\biggr)
\exp\biggl(\sum_{n=1}^{\infty}\frac{b_{n}}{n}z^{-n}\biggr)\o \e^{-1}
z^{-\d_\e},
\tag 1.13\endgather
$$
Fourier components of these series define linear operators on $H$.

\proclaim{Theorem 1.3} (Boson-fermion correspondence; \cite{F})
The series $\psi,\psi^*$ satisfy the fermionic commutation relations
$$
\gather
\psi(z)\psi(w)+\psi(w)\psi(z)=
\psi^*(z)\psi^*(w)+\psi^*(w)\psi^*(z)=
0,\\ \psi^*(z)\psi(w)+\psi(w)\psi^*(z)=
\delta(z-w)=\sum_{n\in\Z}z^nw^{-n}.\tag 1.14
\endgather
$$
In particular, the operators $\psi_n,\psi^*_n$ satisfy the relations of the
Clifford
algebra, i.e.
$$
\gather
\psi_m\psi_n+\psi_n\psi_m=
\psi^*_m\psi^*_n+\psi^*_n\psi^*_m=0,\\
\psi^*_n\psi_m+\psi_m\psi^*_n=\delta_{m+n,0}.\tag 1.15\endgather
$$
Furthermore, we have an inverse formula to (1.13):
$$
b_n=\sum_{m\in\Z} \psi_m\psi^*_{n-m}\tag 1.16
$$
(as operators in $H$).
\endproclaim

\vskip .1in

\centerline{\bf 2. Spectral properties of vertex operators}

\vskip .1in

2.1. {\it Vertex operators as power series in $q$.}
Let $\C(q)_0$ be the ring of all rational functions of $q$ smooth at the
point $q=0$. This ring is naturally a subring of the ring of formal
power series $\C[[q]]$, so we have a natural topology on $\C(q)_0$ which
defines the notion of convergence of a Taylor series to a rational function.

Theorems 1.1, 1.2 imply the following important proposition.

\proclaim{Proposition 2.1} The Fourier components
(with respect to $z$) of the operators $\Phi_{\pm}(-q^{-3}z): L\to\hat L$
define $\C(q)_0$-linear endomorphisms of $\C(q)_0\o H$.
More precisely, the operators $\Phi_{\pm}(-q^{-3}z)$
can be written in the form
$$
\Phi_{\pm}(-q^{-3}z)=\sum_{n=0}^{\infty}\Psi_{\pm}^n(z)q^n,\tag 2.1
$$
where $\Psi_{\pm}^n(z)$ are Laurent polynomials in $z$ with coefficients
in $\text{End}(H)$. Furthermore, if
$v\in H$ then every homogeneous component of the series $\Phi_{\pm}(z)v$ is
convergent $q$-adically (as a series with values in a finite rank free
$\C(q)_0$-module).
\endproclaim

Let $H[[q]]$ denote the $\C[[q]]$-module consisting of all formal series
$w=\sum_{n\ge 0}w_nq^n, w_n\in H$. Then we have

\proclaim{Corollary 2.2} For any complex number $z\in\C^*$, the operators
$\Phi_{\pm}(-q^{-3}z)$ define $\C[[q]]$-endomorphisms of $H[[q]]$.
\endproclaim

 From now on vertex operators will be regarded as such endomorphisms.

\vskip .1in

2.2. {\it Composition of vertex operators.}

Proposition 1.3 implies that we can define composition of
any number of vertex operators, as a formal series in $q$. In particular,
we can define
$$
F_{\e_1\e_2...\e_n}(q)=\Phi_{\e_n}(-q^{-3})...\Phi_{\e_2}(-q^{-3})
\Phi_{\e_1}(-q^{-3}),\
\e_i\in\{+,-\}\tag 2.2
$$
(this was first shown in \cite{DJO}).
We will be especially interested in the operators
$F_{-+}(q)=\Phi_+(-q^{-3})\Phi_-(-q^{-3})$ and $F_{+-}(q)=
\Phi_-(-q^{-3})\Phi_+(-q^{-3})$,
in particular, their leading eigenvectors and eigenvalues.

{\it Remark. } The operator
$F$ is defined over $\C[[q]]$ but, in general, not over $\C(q)_0$
(if it contains two factors or more). Indeed, according to \cite{JM},
the diagonal matrix element of $F_{-+}$ corresponding to the vacuum vector in
$L_0$
equals $\frac{(q^6;q^4)_{\infty}}{(q^4;q^4)_{\infty}}$, where $(a,p)_{\infty}$
denotes $\prod_{n=0}^\infty (1-ap^n)$. This function is obviously not rational,
but it is defined as an element of $\C[[q]]$.

\vskip .1in

2.3. {\it The operators $F_{+-}(0)$ and $F_{-+}(0)$.}

Let us denote by $H_n$ the subspace of $H$ spanned by all the
vectors $P\o\e^n$, $P\in \C[b_{-1},
b_{-2},...]$. Clearly, the operators $F_{+-}(q)$, $F_{-+}(q)$ preserve the
space $H_n$
for all $n\in\Z$.

\proclaim{Proposition 2.3}

(i) The operator $F_{-+}(0)$ preserves degree in $H_0$.
It satisfies the equation $F_{-+}(0)v_0=v_0$ and is nilpotent in
homogeneous subspaces of strictly negative degree in $H_0$.
In $H_n$, the operator $F_{-+}(0)$ lowers the degree by $n$.

(ii) The operator $F_{+-}(0)$ preserves degree in $H_{1}$.
It satisfies the equation $F_{-+}(0)v_{1}=v_{1}$
and is nilpotent in homogeneous subspaces of
strictly negative degree in $H_{1}$.
In $H_n$, the operator $F_{+-}(0)$ lowers the degree by $n-1$.

\endproclaim

The rest of Section 2.3 is the proof of this proposition.
Since (ii) is analogous to (i), we prove only (i).

Substituting $q=0$ in (1.10)-(1.12), we get
$$
\gather
F_{-+}(0)=\phi^*_0, \text{ where }\sum_{n\in\Z}\phi^*_nz^{-n}=\\
\exp\biggl(-\sum_{n=1}^{\infty}\frac{b_{-n}}{n}z^n\biggr)
\exp\biggl(\sum_{n=1}^{\infty}\frac{b_{n}}{n}z^{-n}\biggr)(1\o z^{-\d_\e}),
\tag 2.3\endgather
$$
{}From this formula, it is obvious
that $F_{-+}(0)$ lowers degree by $n$ in $H_n$, in particular,
preserves degree in $H_0$, and that it fixes the vector $v_0$.
It remains to prove the nilpotency of this operator
 on vectors of negative degree.

\proclaim{Lemma 2.4} The operators $\phi^*_n$ in $H$ satisfy the quadratic
relations
$\phi^*_n\phi^*_{m-1}+\phi^*_m\phi^*_{n-1}=0$.
\endproclaim

\demo{Proof} This Lemma follows from the boson-fermion correspondence
(Theorem 1.3).
Indeed, we see that $\phi^*(z)=\psi^*(z)(1\o z^{\d_\e}\e^1)$.
This means that
$$
\phi^*(z)\phi^*(w)|_{H_0}=w(1\o\e)\psi^*(z)\psi^*(w)(1\o \e)|_{H_0},\tag 2.4
$$
which implies $w^{-1}\phi^*(z)\phi^*(w)+z^{-1}\phi^*(w)\phi^*(z)=0$ in $H_0$.
This is equivalent
to the identities $\phi^*_n\phi^*_{m-1}+\phi^*_m\phi^*_{n-1}=0$.
$\square$\enddemo

In particular, Lemma 2.4 implies that $\phi^*_0\phi^*_{-1}=0$ in $H_0$.
Similarly, $(\phi^*_0)^2\phi^*_{-2}=-\phi^*_0(\phi^*_{-1})^2=0$. Continuing
this,
by induction we obtain $(\phi^*_0)^k\phi_{-k}=0$. Therefore, the nilpotency
in Proposition 2.3
follows from the following Lemma.

\proclaim{Lemma 2.5} The vectors
$(\phi_{-k}^*)^{n_k}\dots(\phi_{-2}^*)^{n_2}(\phi_{-1}^*)^{n_1}v_0$, where
$k,n_1,...,n_k$ are
any nonnegative integers, form a basis in $H_0$.
\endproclaim

\demo{Proof} Let $H_0'$ be the space spanned by the vectors from Lemma 2.5.
Note that there are exactly as many vectors of each degree among them
as the dimension of the corresponding homogeneous subspace in $H_0$. So,
 in order to prove the Lemma, it suffices to show that $H_0'=H_0$.

In order to establish this, let us first show that $H_0'$ is invariant under
the operators $\phi^*_n$, $n\in\Z$. Indeed, using relations from Lemma 2.4,
we can rearrange factors in any monomial of $\phi^*_n$-s so that the
subscripts increase from left to right. But such a monomial reduces to a
monomial with only negative indices, since $\phi^*_nv_0=0$, $n>0$, and
$\phi^*_0v_0=v_0$. This implies that $\phi^*_n$ maps $H_0'$
to itself for any $n$.

Now let us introduce a new series
$$
\phi(z)=\psi(z)(1\o z^{-\d_\e-1}\e^{-1})=
\exp\biggl(\sum_{n=1}^{\infty}\frac{b_{-n}}{n}z^n\biggr)
\exp\biggl(-\sum_{n=1}^{\infty}\frac{b_{n}}{n}z^{-n}\biggr).\tag 2.5
$$
Similarly to Lemma 2.4, we can prove the relations
$$
\phi_n\phi_{m+1}+\phi_m\phi_{n+1}=0,
\phi_m\phi^*_n+\phi^*_n\phi_m=\delta_{m+n,0}.\tag 2.6
$$
We also have $\phi_0v_0=v_0$, as follows from (2.5).

Let us show that the operators $\phi_n$ leave $H_0'$ invariant.
For $n\ge 0$, this is obvious because of (2.6). In the case $n<0$, it is
enough to prove that $\phi_nv_0\in H_0'$.

Consider the series
$$
u(s_1,...,s_m,z)=\phi^*(s_1z)\phi^*(s_2z)\dots \phi^*(s_mz)v_0=
\prod_{i<j}(1-s_j/s_i)\exp(-\sum_{n=1}^\infty \frac{b_{-n}}{n}(\sum_js_j^n)z^n)
v_0.\tag 2.7
$$
Let $u(s_1,...,s_m,z)=\sum_{n\ge 0}u_{-n}(s_1,...,s_m)z^n$. It is clear
that $u_{-n}\in H_0$ for any numbers $s_1,...,s_m$ such that
$|s_1|>|s_2|>...>|s_m|$ (since it is a sum of a convergent
series of homogeneous vectors in $H_0$). By analytic continuation
$u_{-n}\in H_0$ for any nonzero values of $s_1,...,s_m$.
In particular, setting $s_k=e^{2\pi i(k-1)/m}$, we get
$u_0=v_0$, $u_{-1}=u_{-2}=...=u_{-m+1}=0$, $u_{-m}=Cb_{-m}v_0$, where
$C$ is a nonzero constant. We conclude that $b_{-m}v_0\in H_0'$.
But due to (1.16)
we have $b_{-m}v_0=\phi_{-m}v_0+\phi_{-m+1}\psi_{-1}^*v_0+...+
\phi_0\phi_{-m}^*v_0$. By induction in $m$, we get that $\phi_{-m}v_0\in H_0'$,
i.e $H_0'$ is invariant under $\phi_{-m}$.

Because of (1.16), this implies that $H_0'$ is invariant under $b_{-m}$,
$m>0$, i.e. $H_0'=H_0$, Q.E.D.
$\square$\enddemo

Proposition 2.3 is proved. $\square$

{\it Remark. } The connection between the $q\to 0$ limit of the
vertex operator construction of level one $\U$-modules and the boson-fermion
correspondence which was utilized in our proof was found by I.Frenkel
and N.Jing (private communication).

\vskip .1in

2.4. {\it The highest eigenvalue of $F_{-+}(q)$, $F_{+-}(q)$.}

\proclaim{Proposition 2.6} (i) There exists a unique
vector $u_0(q)=v_0+u_0^1q+...\in H[[q]]$
such that its zero degree component is $v_0$,
and a unique formal series $\l(q)=1+\l_1q+...\in\C[[q]]$
such that $F_{-+}(q)u_{0}(q)=\l(q)u_0(q)$.

(ii) There exists a unique $F_{-+}(q)$-invariant $\C[[q]]$-submodule $U_0$
in $H[[q]]$ such that $H[[q]]=\C[[q]]u_0(q)\oplus U_0$.

(iii)
There exists a unique
vector $u_{1}(q)=v_{1}+u_{1}^0q+...
\in H[[q]]$ such that its zero degree component is $v_1$, and a unique
formal series $\l^*(q)=1+\l_1^*q+...\in\C[[q]]$
such that $F_{+-}(q)u_{1}(q)=\l^*(q)u_{1}(q)$.
The series $\l^*$ coincides with $\l$.

(iv) There exists a unique $F_{+-}(q)$-invariant $\C[[q]]$-submodule $U_{1}$
in $H[[q]]$ such that $H[[q]]=\C[[q]]u_{1}(q)\oplus U_{1}$.
\endproclaim

\demo{Proof} Since (iii), (iv) are analogous to (i), (ii), we prove (i),
(ii) only.

(i) Let
$$
F_{-+}(q)=\sum_{n\ge 0}F_nq^n.\tag 2.8
$$
Let us look for $u_0,\l$ in the form
$$
u_0(q)=\sum_{n\ge 0} u_0^nq^n, \l(q)=\sum_{n\ge 0} \l_nq^n, \l_0=1,
 u_0^0=v_0.\tag
2.9
$$
Then from $F_{-+}u_0=\l u_0$ we get
$$
\sum_{m=0}^n F_mu_0^{n-m}=\sum_{m=0}^n\l_mu_0^{n-m},\ n\ge 0.\tag 2.10
$$
This can be rewritten as a recursive relation
$$
(F_0-1)u_0^n=\l_nv_0-F_nv_0+\sum_{m=1}^{n-1}(\l_{n-m}-F_{n-m})u_0^m.\tag
2.11
$$
This implies, in particular, that all vectors $u_0^n$ must belong to $H_0$.

The operator $F_0-1$ is not invertible (it kills $v_0$), but it is
invertible on vectors of negative degree in $H_0$, by virtue of
Proposition 2.3. Therefore, we must choose $\l_n$ in such a way that
the right hand side of (2.11) does not have a zero degree term.
This can be done in a unique way. After $\l_n$ is chosen,
$u_0^n$ is determined uniquely by
$$
u_0^n=(F_0-1)^{-1}(\l_nv_0-F_nv_0+\sum_{m=1}^{n-1}(\l_{n-m}-F_{n-m})u_0^m).
\tag 2.12
$$
(because $u_0^n$ has to have a trivial zero degree component).

(ii) To define an invariant
 complement $U_0$ to the eigenvector $u_0$ is the same as
to define a $\C[[q]]$-linear function $\theta: H[[q]]\to\C[[q]]$
such that $F_{-+}^*\theta=\l\theta$ and $\theta(u_0)=1$
($\theta$ is the projection along $U_0$, $U_0$ is the kernel of $\theta$).
It is shown in the same way as
in the proof of (i) that such a function is unique.
$\square$\enddemo

\vskip .1in

\centerline{\bf 3. The semi-infinite tensor product construction.}

\vskip .1in

3.1. {\it The Kyoto conjecture.}

Consider the matrix elements
$$
G^0_n(q)=\<v_0^*,F_{-+}(q)^nv_0\>,
\ G^{1}_n(q)=\<v_{1}^*,F_{+-}(q)^nv_{1}\>
\tag 3.1
$$
where $v_i^*$ are the lowest weight vectors in $L_i^*$ such that
$\<v_i,v_i^*\>=1$.

Clearly, $G^i_n(q)\in\C[[q]]$.

The following statement was conjectured in \cite{DFJMN}
(we call it ``the Kyoto conjecture'').

\proclaim{Theorem 3.1} The sequence $G^i_n(q)^{1/n}$ for $i=0$ or $1$ is
$q$-adically convergent,
and its limit equals $\l(q)$.
\endproclaim

\demo{Proof} We give the proof in the case $i=0$. The case $i=1$ is analogous.

Let us write $v_0$ in the form $v_0=\xi(q)u_0(q)+w(q)$,
where $\xi\in\C[[q]]$, $w\in U_0$.
This can be done in a unique way. Then by Proposition 2.6 we have
$$
F_{-+}(q)^nv_0=\xi(q)\l(q)^nu_0(q)+F_{-+}(q)^nw(q).\tag 3.2
$$
So, it is enough to show that for any $N>0$ $F_{-+}(q)^nw(q)$ is zero
in $U_0/q^NU_0$ for a sufficently large $n$.
That is, to show that $F_{-+}(q)$ is locally nilpotent in $U_0/q^NU_0$.

Let $W$ be the subspace of $H_0$ spanned by all vectors of strictly negative
degree. Let $P:U_0\to W[[q]]$ be the projection parallel to $v_0$.
Let $M(q)=PF_{-+}(q)P^{-1}: W[[q]]\to W[[q]]$. Then
$M(q)=\sum_{n\ge 0}M_nq^n$, $M_n\in\text{End}W$, and $M_0=F_0|_W$.
It is enough to prove local nilpotency of $M(q)$ in $W[[q]]/q^NW[[q]]$.

Fix $N$. We have $M(q)=\sum_{n=0}^{N-1}M_nq^n$ in $W[[q]]/q^NW[[q]]$.
Let $w\in W$ be a homogeneous vector
of degree $m$. Let $d_n$ be the smallest degree of a nontrivial
homogeneous component of $M_n$ (remember that this degree is nonpositive).
Let $d^*=\min_nd_n$. Let $r$ be a positive integer such that
$F_0^{r+1}=0$ on vectors in $H_0$ of degree $\ge m+(N-1)d^*$.
Such $r$ exists because of Proposition 2.3.

Then $M(q)^{Nr+N}w=0$ in $W[[q]]/q^NW[[q]]$. Indeed,
let us expand the power of $M(q)$. Then any term contributing to the
coefficient to $q^k$, $k\le N-1$, will look like
$F_0^{r_1}M_{s_1}F_0^{r_2}M_{s_2}...F_0^{r_l}M_{s_l}F_0^{r_{l+1}}w$,
where $l\le k$. Since $l+\sum_{j=1}^{l+1} r_j=Nr+N$,
we have that at least one $r_j$ is
$\ge r+1$. Since the degree of any homogeneous component of the vector
to which $F_0^{r_j}$ is applied in our term is clearly $\ge m+(N-1)d^*$
(remember that $F_0$ preserves degree), it follows from the choice of $r$
that the whole term is zero.
$\square$\enddemo

{\it Remark. } In \cite{DFJMN}, the authors use the operators
$\Phi(1)$ rather than $\Phi(-q^{-3})$. However, this variation
does not affect quantity
(3.1), so all our arguments remain valid.

Actually, our method of proof of Theorem 3.1 allows to prove a more general
statement, also conjectured in \cite{DFJMN}.

\proclaim{Theorem 3.2}(i) Let $w\in H$. Then there exist formal limits
$$
\eta_0(w)=\lim_{n\to\infty}\l(q)^{-n}\<v_0^*,F_{-+}(q)^nw\>,\
\eta_1(w)=\lim_{n\to\infty}\l(q)^{-n}\<v_1^*,F_{+-}(q)^nw\>.\tag 3.3
$$

(ii) $\eta_i(w)=\theta_i(w)\<v_i^*,u_i\>$, where $u_0,u_1$ are the eigenvectors
of the operators $F_{-+}$, $F_{+-}$, and $\theta_0,\theta_1$
are the linear functionals defined by $w\in\theta_i(w)u_i+U_i$, $i=0,1$.
\endproclaim

\demo{Proof} Analogous to Theorem 3.1.
\enddemo

\vskip .1in

3.2. {\it The Kyoto homomorphism.}

Let $S$ be the set of sequences
$\{p_n,n\ge 1\}$, $p_n\in\{+,-\}$, such that there exists $N=N(p)$
such that for $n>N$ $p_n=-p_{n-1}$. An element $p\in S$ is called a path.
A path $p$ is called odd if $p_n=(-1)^{n-1}$ for sufficiently
large $n$, and even if $p_n=(-1)^n$ for sufficiently large $n$.
The set of odd (even) paths is denoted by $S_1$ (respectively $S_0$),
so $S=S_0\cup S_1$. Let $T_i=\C[S_i]$, $i=0,1$, and $T=\C[S]=T_0\oplus T_1$
be the spaces of functions on $S_i,S$ which vanish almost everywhere.
One can interpret $T$ as a semiinfinite tensor product
$...\o V\o V$, where $V=\C v_+\oplus \C v_-$ is a 2-dimensional
representation of $\U$.
let $T^*[[q]], T^*((q))$ be the sets of all linear maps from $T$ to
$\C[[q]],\C((q))$ (here $\C((q))$ is the field of formal Laurent series).
Following \cite{DFJMN}, let us define a $\C(q)$-linear map
$K: L\to T^*((q))$, as follows.

\proclaim{Definition} The Kyoto homomorphism is the linear map
$K: L\to T^*((q))$ defined by
$$
(Kw)(p)=\l(q)^{-n}\eta_i(\Phi_{p_{2n}}(1)...\Phi_{p_1}(1)w)=
=\l(q)^{-n}\eta_i(F_{p_1p_2...p_{2n}}(-q)^{3d}w), w\in L,\ p\in S_i,\tag
3.4
$$
where $n$ is any positive integer for which $p_{N+1}=-p_N$, $N\ge 2n$, and $d$
is the operator of homogeneous degree.
\endproclaim

\proclaim{Lemma 3.3} The map $K$ is well defined, i.e. does not depend on
the choice of $n$.
\endproclaim

\demo{Proof} Let $n,m$ be two positive integers satisfying the conditions
of the definition, and let $n<m$. Then they give the same value of $K$ because
of the identity $\eta_0(F_{-+}w)=\l(q)\eta_0(w)$,
 $\eta_1(F_{+-}w)=\l(q)\eta_1(w)$.
$\square$\enddemo

It is clear that the map $K$ sends $L_i$ into
$T_i^*((q))$, $i=0,1$.

In particular, we can define the ``half-vacuum state''
$s_0(q)=Kv_0$. Clearly, $s_0\in T_0^*[[q]]$.

\proclaim{Proposition 3.4} $(s_0(0))(p)$ equals $1$
 if $p_n=(-1)^n$, $n\ge 1$, and $0$ otherwise.
\endproclaim

\demo{Proof} It is easy to check that at $q=0$ the product
$\Phi_{\e_{2n}}(-q^{-3})....\Phi_{\e_1}(-q^{-3})$ vanishes whenever
$\e_j=\e_{j+1}=+$ for some $j$. This is easy to show using the relation
$\phi_m^*\phi_{m-1}^*=0$. Therefore, if the product
 $\Phi_{\e_{2n}}(-q^{-3})....\Phi_{\e_1}(-q^{-3})v_0$
does not vanish at $q=0$, then $\e_1=-$ and two pluses cannot stand beside
each other. So, if in addition the total numbers of pluses and minuses
are the same (i.e the considered vector is in $H_0[[q]]$) then
the only possibility is $\e_j=(-1)^j$. This implies the proposition.
$\square$\enddemo

{\it Remark.} The main part of the conjecture in \cite{DFJMN} is to show
that when the map $K$ (whose very existense was so far conjectural)
is applied to the vector $G(p)$ of Kashiwara's upper global base of $L$
corresponding to the path $p$, then the obtained functional in $T^*((q))$
($KG(p)$) is actially in $T^*[[q]]$, and tends to the characteristic
function of $p$ at $q\to 0$. The above arguments do not settle this
question, at least without some additional work; one needs a certain
technique of keeping track of leading degrees of $q$.
We will discuss it in a later paper.

\vskip .1in

\centerline{\bf 4. Computation of the leading eigenvalue and eigenvector.}
\vskip .1in

It turns out that the eigenvalue $\l(q)$ and the eigenvectors
$u_0(q),u_1(q)$ of $F_{-+}(q),F_{+-}(q)$ can be computed explicitly.
The following theorem was recently proved by M.Jimbo.

\proclaim{Theorem 4.1} The following identities hold:
$$
\lambda(q)=\frac{(q^6;q^8)^2_\infty}{(q^4;q^8)^2_\infty},\tag 4.1
$$
where $(a,p)_{\infty}$
denotes $\prod_{n=0}^\infty (1-ap^n)$;
$$
u_0(q)=e^{\Cal F_0}v_0,\ u_1(q)=e^{\Cal F_1}v_1,\tag 4.2
$$
where
$$
\Cal F_0=-\frac{1}{ 2}\sum_{n=1}^\infty\frac{1+q^{2n}}{n}q^{2n}b_{-n}^2
-\sum_{n=1}^\infty \frac{1}{n}(-q)^{3n} b_{-n}
-\sum_{n=1}^\infty \frac{1-q^{2n}}{2n}q^{2n} b_{-2n},\tag 4.3
$$
$$
\Cal F_1=-\frac{1}{ 2}\sum_{n=1}^\infty\frac{1+q^{2n}}{n}q^{2n}b_{-n}^2
+\sum_{n=1}^\infty\frac{(-q)^n}{n}b_{-n}
-\sum_{n=1}^\infty \frac{1-q^{2n}}{2n}q^{2n} b_{-2n}\tag 4.4
$$
\endproclaim

In particular, the series
$\l(q)$ defines a nonvanishing analytic function in the region $|q|<1$.

{\it Remark. } We see that $\l(q)^{1/2}=1+q^4-q^6+q^8 \text{ mod }q^{10}$,
which was found in \cite{DFJMN}.

\demo{Proof} Let us first prove the formula for $u_0$.
In the sequel we assume that $q$ is a complex number with $|q|<1$.
For brevity we will write $\Phi_-$ instead of $\Phi_-(-q^{-3})$,
and $\sum$ for $\sum_{n=1}^{\infty}$. The index $n$ will always take
positive integer values.

According to Section 1,
we have
$$
F_{-+}=\frac{1}{2\pi i}\int_{|z|=1}Y(z)\frac{dz}{z},
\ Y(z)=z(\Phi_-X^-(z)-qX^-(z)\Phi_-)\Phi_-
\tag 4.5
$$
(the contour is oriented anticlockwise).
Substituting (1.10) and (1.12) into (4.5), after normal ordering
(i.e. putting terms with $b_{-n}$ to the right, and with $b_n$ to the left)
we obtain
$$
\gather
Y(z)|_{H_0}=\frac{1-q^2}{(1+qz^{-1})^2(1+qz)}\frac{(q^2;q^4)_{\infty}}
{(q^4;q^4)_\infty}\times\\
\exp\biggl(\sum (2(-q)^n-(1+q^{2n})z^n)\frac{b_{-n}}{n}
\biggr)\exp\biggl(-\sum (\frac{2(-q)^n}{1+q^{2n}}-z^{-n})\frac{b_n}{n}\biggr).
\tag 4.6\endgather
$$
Let us look for an eigenvector of $F_{-+}$ in $H_0$ in the form
of an exponential function of a quadratic polynomial:
$$
u_0=\exp(\sum (\beta_nb_{-n}^2+\gamma_n b_{-n}))v_0,\tag 4.7
$$
where $\beta_n,\gamma_n$ are undetermined coefficients depending on $q$.
Applying $F_{-+}$ to (4.7) and using (4.5),(4,6), after normal ordering we get
$$
\gather
F_{-+}u_0=\frac{1}{2\pi i}\exp(\sum (\beta_n b_{-n}^2+\gamma_n b_{-n}))
\exp\biggl(\sum \frac{2(-q)^n}{n}
(1-\frac{2\beta_nn}{1+q^{2n}})b_{-n}\biggr)\times\\
\int_{|z|=1}
\exp\biggl(\sum (2\beta_nz^{-n}-\frac{(1+q^{2n})z^n}{n})b_{-n}\biggr)
g(z)\frac{dz}{z},\tag 4.8\endgather
$$
where
$$
\gather
g(z)=\frac{1-q^2}{(1+qz^{-1})^2(1+qz)}\frac{(q^2;q^4)_{\infty}}
{(q^4;q^4)_\infty}h(z)\\
h(z)=\exp\biggl(\sum \beta_n(z^{-n}-\frac{2(-q)^n}{1+q^{2n}})^2+\sum\gamma_n
(z^{-n}-\frac{2(-q)^n}{1+q^{2n}})\biggr).\tag 4.9\endgather
$$
Therefore,
the identity $F_{-+}u_0=\l u_0$ that we would like to satisfy can be rewritten
in the form
$$
\gather
\frac{1}{2\pi i}
\int_{|z|=1}\exp\biggl(\sum (2\beta_nz^{-n}-\frac{(1+q^{2n})z^n}{n})b_{-n}
\biggr)
g(z)\frac{dz}{z}=\\
\l(q)\exp\biggl(-\sum \frac{2(-q)^n}{n}
(1-\frac{2\beta_nn}{1+q^{2n}})b_{-n}\biggr).\tag 4.10\endgather
$$
Let us compute
the integral on the l.h.s. of (4.10).
Assume that $h(z)$, as a power series in $z^{-1}$, defines a function
holomorphic in the region
$|z|\ge |q|\delta$ for some $\delta<1$
 (including $z=\infty$), and $h(-q)=0$, $h'(-q)\ne 0$.
(We will later check that these conditions are satisfied for the undetermined
coefficients we are going to choose). Then the function $g(z)$ has a simple
pole at $z=-q$. Therefore, by the residue formula, the l.h.s. of (4.10) equals
$$
\gather
\frac{1}{2\pi i}
\int_{|z|=|q|\delta}
\exp\biggl(\sum (2\beta_nz^{-n}-\frac{(1+q^{2n})z^n}{n})b_{-n}\biggr)
g(z)\frac{dz}{z}\\
+\exp\biggl(\sum (2\beta_n(-q)^{-n}-\frac{(1+q^{2n})(-q)^{n}}{n})b_{-n}\biggr)
\lim_{z\to -q}g(z)(1+qz^{-1})\tag 4.11\endgather
$$
(we have moved the contour of integration through the pole).

We would like the integral term in (4.11)
to be proportional to the integral on the left hand side of (4.10); then we can
express the integral explicitly, via the non-integral term in (4.11).
To see if we can do this, let us consider the change of variable $z\to
q^{2}z^{-1}$ in the
integral term in (4.11) (this will bring us to the contour
$|z|=|q|\delta^{-1}$, which can then be deformed to $|z|=1$,
since there is no singularities between these two contours). We obtain
$$
\frac{1}{2\pi i}
\int_{|z|=1}
\exp\biggl(\sum
(2\beta_nq^{-2n}z^{n}-\frac{(1+q^{2n})q^{2n}z^{-n}}{n})b_{-n}\biggr)
g(q^{2}z^{-1})\frac{dz}{z}.\tag 4.12
$$
Clearly, this integral is proportional to the l.h.s. of (4.10) if
two coditions are satisfied: 1) $2\beta_nq^{-2n}=-\frac{1+q^{2n}}{n}$, and
2) $g(q^{2}z^{-1})=-g(z)$ in the neighborhood of $|z|=|q|$.
So we choose $\beta_n=-\frac{1}{2}\frac{(1+q^{2n})q^{2n}}{n}$ to satisfy
the first condition, and
assume that the second condition holds
(we will later choose the undetermined coefficients $\gamma_n$
 in such a way that it does).
Then we get
$$
\gather
\frac{1}{2\pi i}
\int_{|z|=|q|\delta}
\exp\biggl(\sum (2\beta_nz^{-n}-\frac{(1+q^{2n})z^n}{n})b_{-n}\biggr)
g(z)\frac{dz}{z}=\\
\frac{1}{2}
\exp\biggl(\sum (2\beta_n(-q)^{-n}-\frac{(1+q^{2n})(-q)^{n}}{n})b_{-n}\biggr)
\lim_{z\to -q}g(z)(1+qz^{-1})\\
=\frac{1}{2}
\exp\biggl(-\sum (\frac{2(1+q^{2n})(-q)^{n}}{n})b_{-n}\biggr)
\lim_{z\to -q}g(z)(1+qz^{-1})
.\tag 4.13\endgather
$$
Substituting this into (4.10), we get
$$
\l(q)=\frac{\lim_{z\to -q}g(z)(1+qz^{-1})}{2}.\tag 4.14
$$
Let us now find $h(z)$, i.e. the sequence $\gamma_n$. We have
$$
\frac{g(z)}{g(q^2z^{-1})}=z^2q^{-2}\frac{1+q^3z^{-1}}{1+qz}
\frac{h(z)}{h(q^2z^{-1})}.\tag 4.15
$$
To satisfy property 2, we want this ratio to be equal to $-1$.
This implies that $h(z)$ vanishes at $z=q$. Therefore, it is natural
to look for $h(z)$ in the form
$$
h(z)=(1-q^2z^{-2})\exp(f(z)),\tag 4.16
$$
where $f$ is regular in the region $|z|\ge |q|\delta$. Then from the equation
$g(z)/g(q^2z^{-1})=-1$ and (4.15) we get
$$
\exp(f(z)-f(q^2z^{-1}))=\frac{1+qz}{1+q^3z^{-1}},\tag 4.17
$$
or
$$
f(z)=c-\ln(1+q^3z^{-1}), \ln h(z)=c+\ln(1-q^2z^{-2})-\ln(1+q^3z^{-1}),\tag 4.18
$$
where $c$ depends on $q$. From this equation and (4.9) we get
$$
\gather
-\frac{1}{2}\sum \frac{1+q^{2n}}{n}q^{2n}
(z^{-n}-\frac{2(-q)^n}{1+q^{2n}})^2+\sum\gamma_n
(z^{-n}-\frac{2(-q)^n}{1+q^{2n}})=\\
-\sum \frac{q^{2n}z^{-2n}}{n}+\sum \frac{(-q)^{3n}z^{-n}}{n}.\tag 4.19
\endgather
$$
{}From this it is easy to obtain equations for $\gamma_n$:
$$
\gamma_n=-\frac{(-q)^{3n}}{n}-\frac{1+(-1)^n}{4}\frac{(1-q^{n})q^n}{n}.\tag
4.20
$$
Thus, we have obtained the first formula in (4.2).

It remains to compute $\l(q)$. From (4.14) we get
$$
\gather
\l(q)=-\frac{q}{2}\frac{(q^2;q^4)_{\infty}}
{(q^4;q^4)_\infty}h'(-q)=\frac{(q^2;q^4)_{\infty}}
{(q^4;q^4)_\infty}\exp(f(-q))=\\
\frac{(q^2;q^4)_{\infty}}
{(q^4;q^4)_\infty}\frac{e^c}{1-q^2}=
\frac{(q^6;q^4)_{\infty}}
{(q^4;q^4)_\infty}e^c,\tag 4.21\endgather
$$
where $c$ is defined by (4.18)
So, we need to compute $e^c$. From (4.18) it is seen that $c$ is the free term
in $\ln h(z)$, so from (4.9) we get
$$
\gather
c=\sum\beta_n\frac{4q^{2n}}{(1+q^{2n})^2}-\sum
\gamma_n\frac{2(-q)^n}{1+q^{2n}}=
\sum_n\frac{q^{4n}-q^{6n}}{1+q^{4n}}=\\
\ln\frac{(q^6;q^8)_\infty (q^{8};q^8)_\infty}
{(q^4;q^8)_\infty (q^{10};q^8)_\infty}.\tag 4.22\endgather
$$
Thus, from (4.21) we finally obtain
$$
\l(q)=\frac{(q^6;q^4)_\infty(q^6;q^8)_\infty (q^{8};q^8)_\infty}
{(q^4;q^4)_\infty(q^4;q^8)_\infty (q^{10};q^8)_\infty}=
\frac{(q^6;q^8)^2_\infty}{(q^4;q^8)^2_\infty},\tag 4.23
$$
which proves (4.1).

It is easy to check that our regularity assumptions on the function $h(z)$ hold
true, so the proofs of (4.1) and the first part of (4.2) are complete.

It remains to prove the second part of (4.2).
Now it is immediate. Indeed, it is clear that
$u_1$ is proportional to $\Phi_-u_0$, so after normal ordering in this
expression we get $\Cal F_1=\sum (\beta_n b_{-n}^2+\tilde\gamma_n b_{-n})$
where
$$
\tilde\gamma_n=\gamma_n-2\beta_n\frac{(-q)^n}{1+q^{2n}}
+\frac{(-q)^n}{n},\tag 4.24
$$
which yields formula (4.4) for $\Cal F_1$.
$\square$\enddemo
\vskip .1in

\centerline{\bf Acknowledgements}
\vskip .1in

I would like to thank M.Jimbo for useful remarks and corrections to
the first version of this paper, and for sharing with me the contents
of Section 4. I am also grateful to E.Frenkel, I.Frenkel, D.Kazhdan,
and T.Miwa for useful discussions.

\end